# The executable digital twin: merging the digital and the physics worlds

**H. Van der Auweraer** [1,3], **D. Hartmann** [2]
[1] Siemens Industry Software NV, Strategy & Innovation,
Interleuvenlaan 68, B-3001, Leuven, Belgium
e-mail: **herman.vanderauweraer@siemens.com**

[2] Siemens Industry Software GmbH, Strategy & Innovation,
Otto-Hahn-Ring 6, 81739 Munich, Germany

[3] KU Leuven, Department of Mechanical Engineering,
Celestijnenlaan 300, B-3001, Heverlee, Belgium

## Abstract

While the digital twin has become an intrinsic part of the product creation process, its true power lies in the connectivity of the digital representation with its physical counterpart. Data acquired on the physical asset can validate, update and enrich the digital twin. The knowledge contained in the digital representation brings value to the physical asset itself. When a dedicated encapsulation is extracted from the digital twin to model a specific set of behaviors in a specific context, delivering a stand-alone executable representation, such instantiated and self-contained model is referred to as an Executable Digital Twin. In this contribution, key building blocks such as model order reduction, real-time models, state estimation and co-simulation are reviewed, and a number of characteristic use cases are presented. These include virtual sensing, hybrid testing and hardware-in-the loop, model-based control and model-based diagnostics.

## 1   Introduction

Managing complexity in product design, manufacturing, and operation is a challenge in everyday decision making to provide safe, sustainable, and efficient products and industrial processes. The digital twin, being a virtual "mirror" representation of a real asset, tightly integrating the real and the digital worlds, has become a key enabler to support such decision making for complex systems. It allows informing design, engineering, operational as well as strategic decisions upfront through highly realistic virtual predictions and optimization of their real-world counter parts. As discussed further on in this paper, the digital twin is inclusive, holistic (in the sense of consistently representing all available data) and dynamic as it will continuously be updated along the product lifecycle and enriched with new datasets and models as these become available.

The Executable Digital Twin is then introduced to implement one very specific use case of the digital twin. It is defined as a self-contained, executable, digital representation of a specific behavior of the physical asset that is instantiated from the digital twin for a specific purpose and possibly (but not necessarily) a specific data format (FMI, ...) or runtime environment (cloud, edge, test-bench, embedded, …). It can be calibrated for the individual asset and leveraged by anyone along the asset's lifecycle. The Executable Digital Twin is hence not any longer holistic and dynamic (unless a new version is created or a dedicated, traceable updating process is foreseen) but it becomes a self-contained (and possible tradeable) asset by itself. It however remains linked to the digital twin from which it was derived. The latter aspect distinguishes it from ad-hoc embedded real-time models. Considering this relation, the digital twin paradigm will first be briefly reviewed followed by the definition and discussion of the Executable Digital Twin.

## 2    Digital twin: historic perspective

The principles behind the digital twin are not new. Product and process engineering teams have used 3D renderings of computer-aided design models, asset models and process simulations for decades. NASA was the first to work with mor formal pairing technology from the early days of space exploration to address the challenges to operate, maintain or repair systems away from physical proximity. It uses digital twins to develop new recommendations, roadmaps, and next-generation vehicles and aircraft [1] [2] [3].

The conceptual view that for each physical system a virtual "mirror" can be conceived and realized was first presented by Dr. Michael Grieves around 2002 in a University of Michigan Executive Course on Product Lifecycle Management (PLM) [4]. The actual term digital twin was coined by Grieves [5] and Tuegel [6] in 2011 and further developed in Grieves' consecutive work [7]. The digital twin as a virtual representation of what was manufactured is presented, comparing a digital twin to its engineering design to better understand what was produced versus what was designed.

The concept of digital thread (the use of digital tools and representations for connecting and tracking design, evaluation, and life cycle management) was introduced in [8] in 2013. The Air Force Research Laboratory linked the digital twin to the operational phase of the aircraft lifetime. According to their 2011 vision [6], a digital twin of the individual aircraft –including deviations from the nominal design – should be delivered together with the physical aircraft to be flown virtually through the same flight profiles as recorded for the actual aircraft. Related to this, Tuegel et al. [9] propose to utilize an ultrahigh fidelity model of the individual aircraft to integrate computation of structural responses as a function of the flight conditions, with resulting local damage and material state evolution.

Over the last decade, the concept of the digital twin has been extended, refined but also sometimes redefined to match specific utilization contexts, not uncommonly to fit the objectives of related solution providers. Lately, the research on the digital twin has been expanding from the mere technical use and value creation to more conceptual and formal design methodology related studies. A number of related comprehensive state-of-the-art reviews have been published [10] [11] [12] [13] [14].

The field where the Digital Twin has probably gained the most attention in the past years is in the manufacturing domain where the added value of the combined digital-physical representation matches very well the drive towards cyber-physical systems and smart manufacturing systems [10] [15] [16]. Furthermore, increasingly  the consideration -and use- of the digital twin in the actual operation phase is documented, where obviously the match of data measured from the physical asset combined with the digital twin opens new opportunities for performance optimization, maintenance, prognostics, servicing, etc. [17] [18] eventually leading to integration in asset management [19]. Keeping the digital twin tuned to the evolving physical system during its lifetime (accounting for repair, degradation, behavioral dependency on operational conditions…) is generally claimed as a critical objective but only few concrete examples are documented. The actual use of the digital twin embedded in the operation is an even larger challenge but also opportunity for value creation as will be discussed further on.

## 3    Digital twin characteristics

### 3.1    Digital twin paradigm

Defining the digital twin is a challenge. The concept has been evolving over time and multiple interpretations have been given for -or imposed on- the digital twin depending on a specific context. Formal conceptual modeling efforts are plentiful, including standardization [20] [21], however, the authors tend to consider the digital twin rather as a paradigm and conceptual framework described by its characteristics and getting concrete meaning within an application context [22] [23].

Commonly accepted is the notion that a digital twin is a specific virtual representation of a physical object, being a product, a process plant, an infrastructure system or a production process. The digital twin integrates all data, models, and other information of the physical object generated along its life cycle for a dedicated purpose. The data can be generated during design, engineering, manufacturing, commissioning, operation, or service. Integrating all information is key to leverage existing and create new business opportunities. In all cases however, the objective is to have a digital representation suited to the purpose in terms of level of detail, completeness, accuracy, and execution speed.

This typically enables reproducing the state and behavior of the corresponding system as well as predicting and optimizing its performance. To this purpose, simulation methods and data-based methods are used. The digital twin can be used in product design, simulation, monitoring, optimization and servicing and is an important concept in the industrial Internet of Things (IIoT).

New application fields for the digital twin are continuously being identified, including in healthcare, biosystems including the human body, logistics, agriculture, construction, transport, etc.

## 3.2 Simulation models and data

A key purpose of the digital twin is to analyze, predict and optimize the behavior of the physical object. Simulation- as well as data-based approaches are used hereto [22] [24] [25]. Simulation models are mathematical representations of the object's behavior based on a first principles description of the governing physical laws and adapted to an implementation on a computer system (e.g., through an appropriate discretization). These -parametrized- simulation models can then be executed for the appropriate input conditions and constraints to yield the behavior predictions of interest. Data-driven models use data measured on the physical object itself to predict the future behavior through reduction into black-box models or by synthesizing the object's behavior by applying statistical and/or data analytics techniques. Ideally both approaches complement each other.

Simulation models can already be applied when no physical system is available yet and allow parametric design optimization, enabling the correspondingly realized physical system to optimally match the required performance. Typical simulation model approaches in the mechanical design space include geometry-based Finite Element models, Boundary Element models, Multibody Dynamics models, Computational Fluid Dynamics models and lumped-parameter System Simulation models which can e.g. be formulated as Ordinary Differential Equations (ODE) or Differential Algebraic Equations (DAE) [22]. While extremely powerful to get insight in the detailed product behavior and to run large numbers of "virtual" design variants to explore the design space, simulation approaches will always suffer to a lesser or larger extent from incompleteness in the modeling approaches, inaccurate parameters, and methodological approximations. They also face the challenge that the manufacturing and operating conditions may affect the actual system as used in its operational environment. From the early days of the digital twin, large emphasis was hence put on developing high fidelity simulation models, enabling to characterize the individualized behavior of the "as manufactured" and "as operated" product.

Data driven models use the data measured on the physical system and as such accurately represent the behavior as observed. Data reduction, advanced analysis, black-box modeling, and data analytics approaches allow to not only get insight in the momentary and historic performance but to a certain degree to also make performance predictions within the envelope of the observed data. The extrapolation to unmeasured phenomena, attributes, operating conditions is however not possible.

While not engaging into the discussion whether a digital twin formally requires the explicit use of measured data (or not) as it may easily lead to conceptual discussions as for Theseus' ship, it is obvious that the combination of the rich information of the simulation models with the actual data as measured on the physical system offers the best potential for analysis, prediction, and performance optimization [26]. This can be in the form of feedback from test data to validate and update and/or complement the simulation models, providing also accurate loading and operating information. For example, this can lead to updating remaining useful life (RUL) models of a structure using actual loading data. But other feedforward scenarios, bringing the simulation model to the level of the data collection, not only for comparison or visualization, are possible and will be discussed further on.

## 3.3 Holistic and dynamic

What makes the digital twin different from a mere simulation model or a collection of test data is that the digital twin concept inherently starts from an integrated view [26]. All simulation models that can be developed, all test data that can be collected, these not all contribute to describing in a digital way as good as possible the physical object's behavior, but these data are inherently connected and refer to the same modeling framework. This requires keeping track of system modifications through cross-referencing and updating all related simulation models and to trace back changing operational conditions to updated performance simulations, staying in tune with each other [27]. This holistic view distinguishes the digital twin from a collection of unrelated simulation models for the various performances of a specific product.

This also implies that the digital twin is not only all encompassing but also dynamic, integrating novel information and novel models as they become available. For example, a system design can start from a functional and system description, to become concrete in a structural (geometric) design and be extended with structural dynamics, thermal, flow, electromagnetic simulation models as these become available. Test data can augment these models with the actual operating information including histories on nominal and non-nominal behavior. Inversely, a digital twin formulation starting from a data-driven model can be enriched with simulation models to increase the predictive power and allow the link to the product design and future design improvement based on actual use information. There seems to be no reason to restrict the digital twin concept to one or the other phase in this process as then the question could be asked from which point onwards the evolving model(s) become(s) a true digital twin.

## 3.4 Empowering the digital twin

As mentioned, the digital twin is an increasingly adopted paradigm with an expanding reach across industrial and societal fields. The confidence of product developers, manufacturers as well as asset operators in digital twin technology is increasing continuously. This trend is strongly enabled by the rapid evolution in several key contributing factors. We highlight four of these, the Industrial Internet of Things, the evolutions on simulation power, the cloud and the use of artificial intelligence (AI).

Undoubtedly, the Industrial Internet of Things with its capabilities for data sensing as well as edge computing, offers enormous opportunities to build and/or enrich the digital twins [28]. For example, in the context of Industry 4.0 the IIoT is a key driver for novel highly automated manufacturing concepts. However, though the IIoT enables data collection on a scale we could not imagine before, data in industrial contexts is still limited. Data is very specific to the context, e.g., the operational data of an electric drive in a car differs significantly from the data in the context of a compressor, and the different contexts addressed in industrial applications cover an enormous breadth. Also, collecting enough data beyond nominal conditions, e.g., for specific failure predictions, is an enormous challenge. The risk of reverse engineering of core industrial process know-how, use conditions etc. furthermore makes industrial companies hesitant to contribute such data outside the company.

Fortunately, many industrial systems are well understood in terms of the first principals or effective engineering models their construction relies on. Computer simulation is used already since several decades as a decision support tool for research, development, and engineering. These digital twins are often trusted to such an extent that even key validation and verification steps are based on them during development and engineering. The exponential evolutions in terms of computer hardware [29], HPC cluster infrastructures, novel hardware solutions like GPU, transputers and in the future possibly quantum computing, open up new worlds of simulation applications. Furthermore, innovative simulation paradigms and algorithm capability [30] [31], including alternate discretization methods, multigrid solvers, meshless methods, etc. have led today to a point where digital twins even allow for interactive simulation and real-time prediction capability thereby opening up novel application opportunities beyond R&D.

One of the hot topics in simulation today is the potential of using AI concepts [32]. This ranges from the use of Machine Learning to (smartly) build up surrogate models which then can be evaluated much faster, to the development of hybrid methodologies integrating different disciplines of mathematical models, algorithms, and machine learning technologies. This opens a potential to challenge many paradigms in computational science and engineering. First success can be seen in the newly emerging fields of scientific machine learning, physics-informed neural networks, or neuro differential equations [33] [34] [35]. Corresponding approaches will, for example, allow the realization of very efficient surrogate models or to complement first-principal models with models based on measurements

Finally, one cannot neglect the transformation in engineering processes and methods realized through the cloud enablement of data collection, simulation, analysis, and engineering. Next to the intrinsic challenges for interoperability, modularization, deployment and SaaS (Software as a Service), new concepts emerge in relation to Digital Twin as a Service [36], model and data traceability (e.g. with blockchain) [37], and (cyber)security [38], while inversely the digital twin can also be used to monitor and enforce security [39].

## 4 The Executable Digital Twin

### 4.1 Scope and Definition

Connecting the physical and the virtual worlds is a key capability of the digital twin. Using data from the physical asset to build the digital twin and/or to integrate with simulation models or even provide use information to the design are obvious tasks. The next logical step, making full use of all capabilities of the digital twin, is however to bring the digital twin itself into the physical world, enabling the knowledge contained in the digital representation to directly create value to the physical asset itself.

To realize this, a dedicated encapsulation is extracted from the digital twin to model a specific behavior (e.g., an individual performance such as strain or vibration and this possibly for a specific part or component such as a gear or bearing) in a specific operational context (e.g. diagnostics). The result is then a dedicated encapsulated and executable representation that can be integrated within the operational execution environment of the physical asset or other virtual environments. The model is not generic anymore and only represents that aspect of the system under study which is relevant for the application. It can be tuned and calibrated for the individual asset it is integrated with.

Such instantiated, self-contained, and packaged model is referred to as the Executable Digital Twin. The abbreviation xDT is becoming broadly accepted.

### 4.2 Applications

Applications for the Executable Digital Twin are multifold. Integrating with sensors on the physical asset, extra, "virtual", sensor signals can be derived augmenting the physical sensor pool and delivering otherwise unmeasurable quantities. The Executable Digital Twin models can also be applied to derive key performance parameters needed for operational performance supervision and diagnostics, maintenance, and prognostics.

The executable models can furthermore be used to represent other hardware systems, components, or parts, in a real-time test environment of the system under test, to allow to execute a virtual system integration and/or control optimization in a Hardware-in-the-loop configuration. When introducing the models in the controllers, model-based control solutions can be developed starting from the same highly accurate original digital twin which is used in the rest of the design engineering process, ensuring full consistency.

The explicit connection that exists between the Executable Digital Twin model and the originating "parent" digital twin allows to maintain consistency and traceability and distinguishes it from the traditional ad-hoc real-time models developed for a specific execution objective as part of the design of the physical asset.

While the above use scenarios may give the impression that the Executable Digital Twin must be deployed on an edge device or other local hardware system connected to a physical asset, this does not preclude the use in a cloud environment. When linking a cloud version of the executable model to the physical asset through IIoT, applications such as remote monitoring, diagnostics, use scenario exploration, and real-time impact assessment become feasible.

Another scenario includes building up system integration models with components delivered from various sources, adhering to a system architecture description. Such scenarios enable virtual product integration for operating condition or performance limit evaluation, requirement validation, procurement and sales support.

In summary, Executable Digital Twins allow leveraging their prediction capability by anyone at any point of a product's life cycle without the need of additional (simulation) software packages [40]. Thus, they allow for the ultimate democratization of the digital twin beyond its creators and creation tools.

## 4.3 Process overview

Three phases can be distinguished in the overall process of the Executable Digital Twin: an authoring, a creation or builder phase and a deployment phase. This is shown in Figure 1.

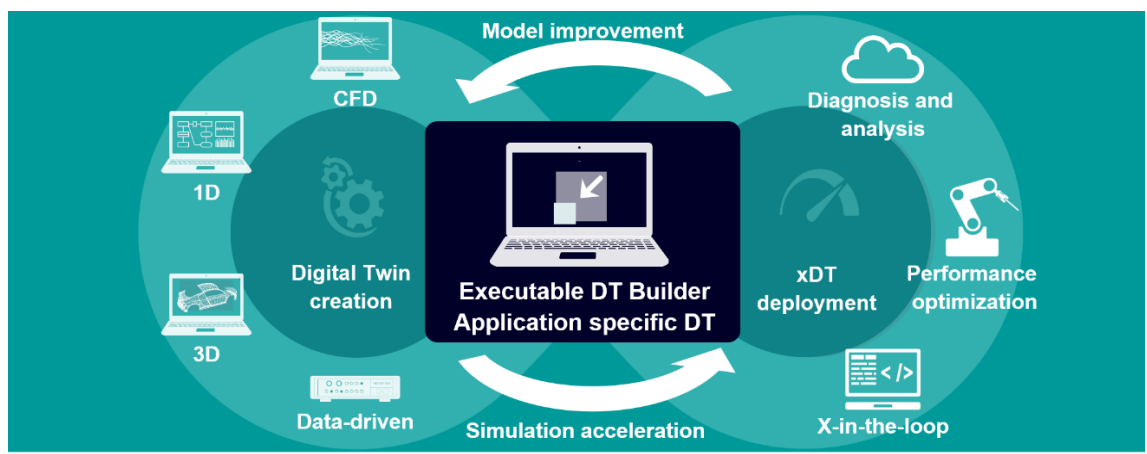

Figure 1: Executable Digital Twin creation and deployment

In the authoring phase, a digital twin is developed containing a variety of interlinked models and data. This digital twin is generic and addresses all aspects of relevance for the physical asset. For a specific purpose and execution context, an application-specific Executable Digital Twin can then be extracted. This results in a containered model for that specific behavior. In the deployment phase, this containered model is picked up and integrated in the application environment (hardware/software). Figure 2 shows the overall flow.

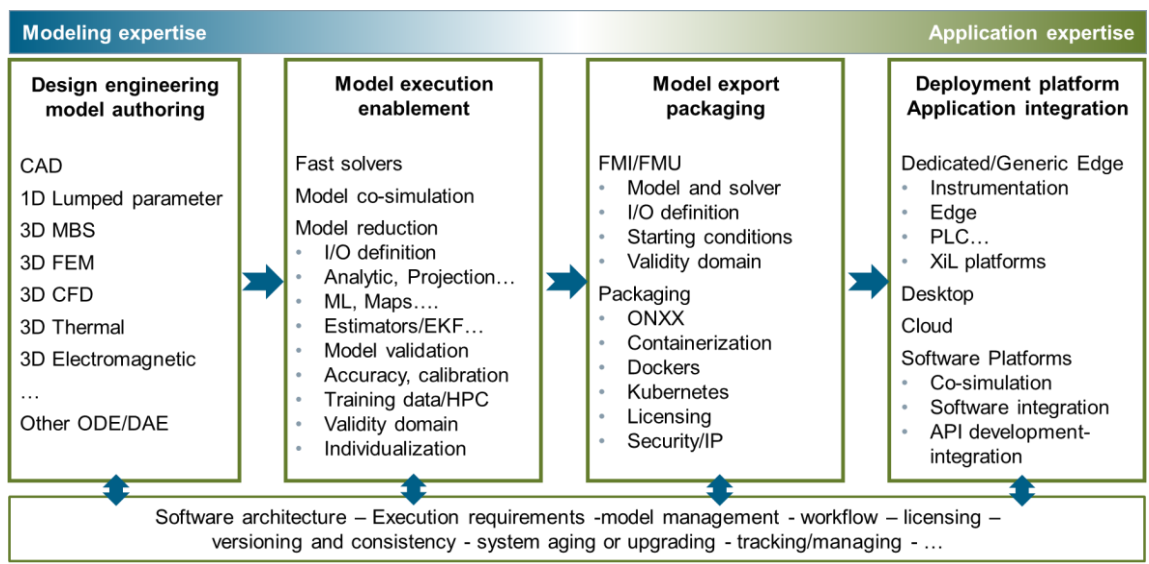

Figure 2: Process flow to author, create and deploy the Executable Digital Twin (xDT)

## 4.4 Enabling Technologies

The Executable Digital Twin relies strongly on a number of key enabling technologies. As real-time (or near-real-time) performance is an essential requirement in most applications, fast simulation methods are required. In most cases this will require the use of powerful Model Order Reduction approaches. State estimation methods can be used to ensure convergence of the virtual sensor signals. As the true value of the Executable Digital Twin is realized in its deployment, hybrid co-simulation schemes accepting packaged models with their execution engines and then linking these in an open runtime environment are key.

### 4.4.1 Model Order Reduction

The central value proposition of the Executable Digital Twin is to provide a "self-contained executable digital behavior of an asset to be leveraged by anyone at any point in the lifecycle". This requires the corresponding model to be sufficiently fast (often real-time), to have an appropriate accuracy (guaranteed or estimated), to be interacted with by a limited set of application programming interfaces (API), to be deployable on various computer hardware.

While the concept of real-time models not new and use cases like virtual sensing or model predictive control are around since long, the new aspect is to allow realizing corresponding use cases in a scalable way. To allow for scalability the concerned Executable Digital Twins are preferably prepared from existing simulation models and historical data that are intrinsically part of the asset's digital twin, i.e., consistency and traceability are maintained contrary to most traditional ad-hoc real-time models.

A key step in doing so, is to programmatically reduce the models to the right model complexity. Models of a limited complexity or fidelity in terms of variables typically still provide predictions with the appropriate accuracy but at less computational effort and at a higher speed of prediction. Today computational science and engineering offers a broad range of technologies. This process is known as model order reduction producing reduced order models with reduced complexity and affordable execution performance.

Generally, corresponding methods can be split into two major categories: Black-box approaches which do not require any information of the underlying system, grey-box approaches which require some knowledge about the associated models, and white-box approaches which require full information and access to the associated models and solvers. A schematic overview of currently applied methods is shown in Figure 3.

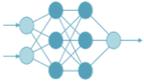

Figure 3: Major model complexity reduction / model order reduction technologies used in the context of Executable Digital Twins.

The abbreviations used refer to the following methods: LTI: Linear Time Invariant; POD: Proper Orthogonal Decomposition; RSM: Response Surface Model; SVD: Singular Value Decomposition; CMS: Component Mode Synthesis; DEIM: Discrete Empirical Interpolation Method; ECSW: Energy Conserving Sampling and Weighing; PGD: Proper Generalized Decomposition; PMOR: Parametric Model Order Reduction

Machine learning models fall in the class of black-box approaches. Typically, corresponding reduced complexity models are trained based on generated simulation data. As the reduced order model is meant to be accessed by a limited set of APIs, the parameter space for which simulations are to be generated is limited. Carefully choosing the parameter sets to be trained on is crucial for success of the applied method [41].

If somethings more is known about the structure of the underlying model, e.g., non-linear relationships, this information can be directly exploited using grey-box models [42] [43]. Instead of taking a neural network representing arbitrary functional relationships the corresponding polynomial class could be used for regression of a model with reduced complexity. Often corresponding approaches need less data than the full neural network models and also show better extrapolation capabilities. Being agnostic to the specific solver black- and grey-box models allow to address solvers where no access to internal information is available.

If access to all relevant solver aspects is possible white box approaches can be used, e.g., Krylov or reduced basis methods [44]. These often allow to optimally tune the reduced order model as well as to provide rigorous accuracy guarantees and thus offer the ultimate model order reduction technology. Whereas multiple approaches exist, e.g., for linear Finite Element Models, adapting the models for time domain predictions and dealing with nonlinear and possibly time variant systems such as in multibody simulation is still a topic of major research [44] [45].

Independent of the specific technology used, all principally allow for a continuous calibration of parameters, enabling ensuring that the Executable Digital Twin is always calibrated with respect to its real counterpart.

### 4.4.2 State Estimation

When the Executable Digital Twin is used to predict virtual measurement values based on a limited set of sensors and an executable model, state estimation techniques are applied [46] [47] [48] [49].

State estimation is a time domain approach to predict the true values of states of a model, given a set of measurements. Two main method families exist: the Kalman filter and the Moving Horizon estimator. The basic idea is to use a weighted sum of both the measurements and model predictions to make an overall optimal prediction of the true value of the states. These states can be internal states to the physical system, or augmented states (states that are added to the system equation, and that can represent inputs/parameters that are unknown or cannot be measured directly). Virtual sensors can be introduced via numerical calculations on top of internal system states, or as augmented states. The choice of weights is often inspired by the measurement noise level and model accuracy. This way, the model is continuously synchronized with measurement data, and can be used for analysis and control.

To achieve optimum accuracy, one must characterize the model prediction errors and the measurement errors by means of their respective covariance matrices. The latter determine the relative weighting of model information against measurement data, i.e., the tuning of the Kalman filter. Because the quantification of the model prediction error covariances is particularly difficult, many applications rely on tedious and sub-optimal manual tuning. Automated procedures are therefore researched [50].

A strongly related issue affecting the quality of the estimated states is the selection of the physical sensors to be used in the estimation process [51] [52] [53].

### 4.4.3 Model Encapsulation

Once the executable model is generated, it is encapsulated and provided to the "consumers" in the deployment environment in an appropriate format.

One of the main approaches used hereto is to consider the Executable Digital Twin as a Functional Mock-up (FMU) that is exchanged in accordance with the Functional Mock-up Interface (FMI) standard. The Functional Mock-up Interface is a free standard [54] that defines a container and an interface to exchange dynamic models. It uses a combination of XML files, binaries and C code zipped into a single file. The FMI standard is supported by many software tools, including the major industrial simulation software providers. Whether the Executable Digital Twin is delivered as a compiled model or as a set of executable equations depends on the application as well as the agreed conventions between the authoring and the user partners. When the Executable Digital Twin is provided by means of machine learning as a neural network, the Open Neural Network Exchange (ONNX) format [55] is sometimes used as an alternative format.

Depending on the application objective and/or environment, container approaches such as Kubernetes [56] and Docker [57] are adopted. In particular, when providing the Executable Digital Twin not as a hardware embedded solution but as a software microservice. Concepts such as Model-as-a-Service (MaaS) [58] may be applied to he Executable Digital Twin and appropriate exchange mechanisms need then to be considered. Finally, compliance specific standards can be required such as prescribed by Autosar [59] when providing the model as an automotive software embedded module.

Future challenges involve the integrity and traceability of the Executable Digital Twin models, connecting directly to the cybersecurity domain [38], including the potential adoption of blockchain concepts [37].

### 4.4.4 Co-simulation

Once the executable model is made available to the deployment environment, it must be integrated in a host application, connecting to the applicable hardware environment in terms of sensors and actuators and be integrated in the host software platform. This may involve executing the model with the solvers as available in the host platform or dedicated solvers are provided together with the model as components in the FMU.

In many cases, the host application will also involve executing simulations of complementary functions, system parts and/or performances. Co-simulation between the Executable Digital Twin model and the host models will then be mandatory. While beyond the scope of this paper and addressed in various surveys [60], it is important to draw the attention to the complexity of co-simulation of potentially heterogeneous model types and systems with different time dependencies, complexity, and behavior (model stiffness) [61].

## 5 Use Cases and Examples

Most of the present industrial use cases documented for the Executable Digital Twin relate to virtual sensing and hybrid testing (combining virtual and physical system components) as executed in the context of the Model Based System Testing company strategy [62] [63]. These applications are still essentially in the realm of the product design and engineering stage with first extensions to the manufacturing and operations field emerging. In this section, several characteristic cases executed in the research teams of the authors are briefly revied and reference is made to the more detailed publications where available. Obviously, a wide range of other applications of the Executable Digital Twin concept are currently explored and realized in other laboratories, often under different headings such as hybrid digital twin and real-time digital twin. The present discussion focuses on own experiences. First an overview of recent virtual sensing cases will be presented, followed by a brief overview of other Executable Digital Twin applications.

### 5.1 Virtual Sensing

Virtual sensing proves to be highly relevant when on-site measurement of the variables of interest is not feasible due to non-accessible locations, sensor cost or the fact that introducing sensors would distort the system under test. In particular, for load measurements, including multi-axial forces and torques, internal temperatures and full field responses, a virtual sensing approach proves to bring large added value.

### 5.1.1 Electric motor startup heating

Large-scale electric drives are exposed to high levels of stress due to induction heating on start-up. Frequent starts without a sufficient interval for cooling can result in motors overheating. However, measuring the temperature of the actual rotors turning at high speed within the motor is close to impossible. Thus, controls are often based on conservative heuristics. The corresponding temperatures can however be calculated by 3D thermal simulations in a sufficiently accurate manner. In the case of the electric motors considered, linear convection-diffusion models are used. Such 3D models are typically available from the detailed engineering phase in product design. Using Krylov model order reduction, corresponding executable models were realized. Continuous calibration is realized through state estimation using available sensors on the stator side. This way, temperatures not accessible to sensors can be virtually measured, see Figure 4.

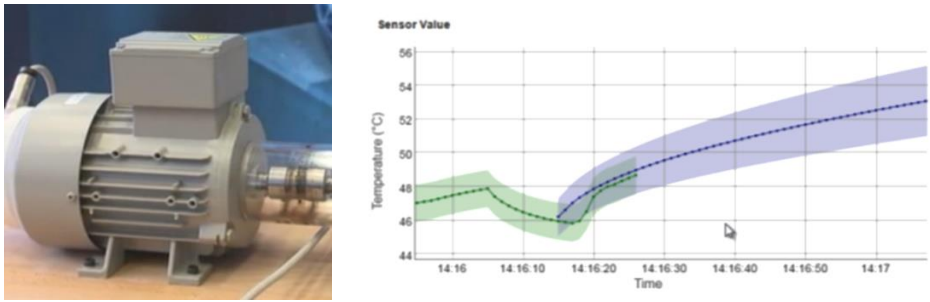

Figure 4: Estimation and uncertainty quantification of unmeasurable rotor temperature (green) and predicting future scenarios of the same temperature (blue). Figure courtesy of Hartman et al. [44]

This allows the cooling times required for electric motors to be significantly reduced, ultimately enhancing the plant availability. Enriched with methods for uncertainty quantification, confidence intervals can be provided for the rotor temperature allowing to go close to operational limits and thus increasing availability. Such an optimized process that can prevent motor overheating and reduces the downtimes required during the cool-off phase, which leads to significant savings per hour. A more detailed discussion is found in [44].

### 5.1.2 Truck anchorage early failure detection

In this second case, smart virtual sensing is used to analyze the cause of failure of a truck anchorage during assembly. Figure 5 shows the setup. The problem was that the failure did not occur during operation but caused a dangerous situation during assembly. Regular virtual testing (FE simulation) alone was not sufficient since the actual loads at the critical process steps are unknown. Some strains are measured but the instrumented strain gauges were not representative for the critical strains at failure.

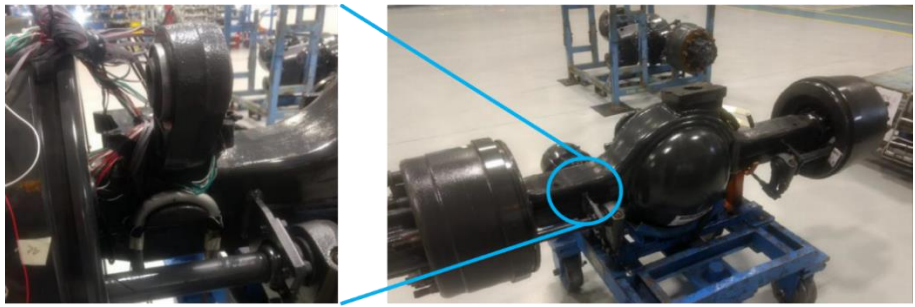

Figure 5: Truck anchorage overview (right) and detail (left). Figure courtesy of Scurria et al. [64]

As the critical locations were inaccessible to measure and only 4 other strain measurements are available and the full strain field under the critical loading conditions is required to analyze the problem, the virtual sensing approach was selected. A linear FE model of the anchorage and the two U-bolts near the critical location was created in Simcenter 3D Nastran. The Simcenter Smart Virtual Sensing tool was then used to create the reduced order model and the corresponding Executable Digital Twin which was then imported in Simcenter Testlab to be fed with the measured strain signals. The Executable Digital Twin then returns the strains at the critical locations and the loads and the full strain field can be estimated. The estimates were validated using a set of independent sensors not included in the virtual sensor. Figure 6 shows the estimated field as well as a comparison of (normalized) measured and estimated strains at the validation locations.

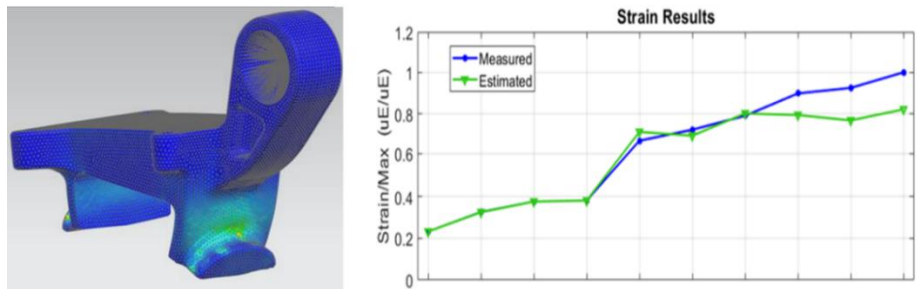

Figure 6: Anchorage full strain field (left) and strain at validation location (right). Figure courtesy of Scurria et al. [64]

The critical locations were accurately predicted, and the estimated strain values turned out to be well in accordance with the measured ones, permitting to rely on the virtual sensing approach for monitoring the assembly process. Correspondingly, a significant improvement in the process of the manufacturer was obtained. A detailed discussion is available in [64].

### 5.1.3   Space payload acoustic test design

To assure that a spacecraft payload can withstand the high dynamic loading during launch, qualification tests need to be performed replicating the high acoustic loading in dedicated test facilities. A novel approach, Direct Field Acoustic eXcitation (DFAX) was developed that allows such tests to be executed by creating the appropriate diffuse acoustic fields using commercially available loudspeakers. Figure 7 shows a representative test setup.

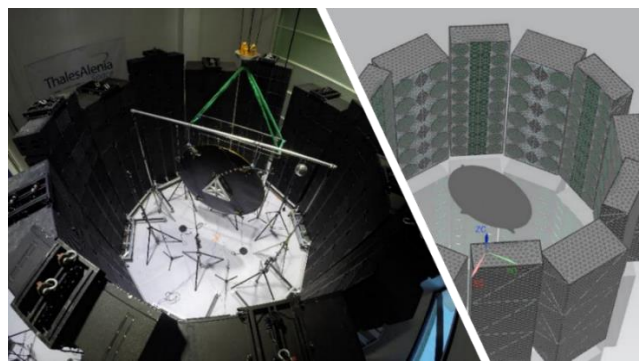

Figure 7: DFAX Test setup, physical twin (left), digital twin (right). Figure courtesy of Thales Alenia Space

A critical step in the process is the generation of the appropriate acoustic field through multiple-input multiple-output control of the loudspeakers. As the system under test should be minimally loaded during the test calibration, predictive virtual testing processes are developed but which need to be tuned on the spot.

Hence a virtual sensing approach was adopted that, starting from a vibro-acoustic model of the test configuration (finite and/or infinite element based) can generate an on-line model that is used in the loudspeaker tuning and diffuse field validation (and where needed payload response verification). In this case, Krylov methods for model reduction were compared with a system identification approach, transforming the numerical model into a frequency response function and subsequently a state-space formulation. Figure 8 shows results of the calibration procedure demonstrating that a valid diffuse field can be generated this way. More details are presented in [65] [66] [67] .

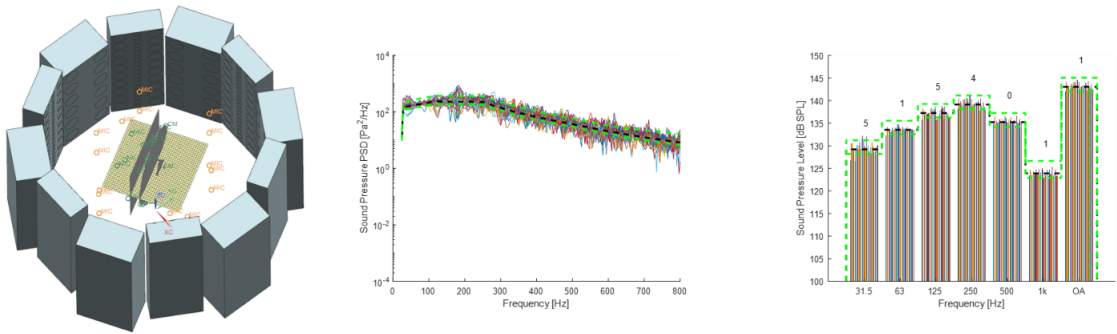

Figure 8: Diffuse field acoustic response validation of the DFAX Executable Digital Twin. Figure courtesy of Alvarez Blanco [65] et al., and van Ophem [66]

### 5.1.4 Wind turbine blade testing

As was already demonstrated in the first case, virtual sensing offers the potential to derive full-field virtual responses using only a limited number of measured sensor signals. This was demonstrated for the case of a wind turbine blade at the Technical University of Denmark (DTU). A composite blade was tested for structural integrity through deflection and dynamic tests, identifying critical design issues but also offering the potential for future in-field Structural Health Monitoring. The blade was instrumented by a limited number of strain sensors only which leads to a long validation process.

A detailed finite element model of the blade was developed and compressed into a reduced order model. Using a state estimator, a virtual sensor for the real-time visualization of the full strain field was realized. Figure 9 shows the test setup and the strain visualization result. A detailed discussion is provided in [68].

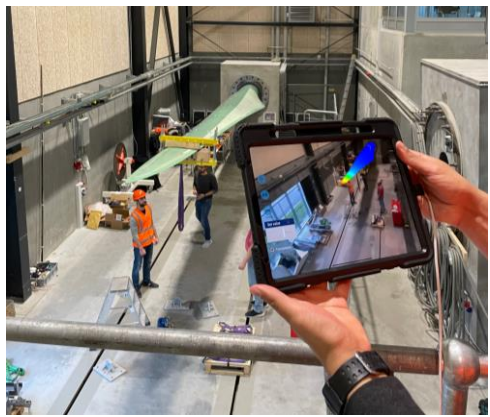

Figure 9: DTU Wind turbine blade test setup and virtual strain field. Figure courtesy DTU and Simcenter Test RTD

### 5.1.5 Vehicle dynamics and durability testing loads and performances

The virtual sensing methodology can also be applied to the estimation of loads and responses for vehicle dynamics benchmarking and validation purposes. In particular, the correct estimation of wheel forces, steering angles and lateral accelerations is of high interest. When enabling the use of only standard available sensors such as the Inertial Measurement Unit (IMU). one can avoid the provision and installation of expensive on-vehicle sensors, including expensive and hard-to-instrument wheel load sensors. In [69] an approach based on a simplified bicycle vehicle model complemented with an adaptive linear tire model is presented, using an Extended Kalman Filtering state estimator. It is shown that not only axle-level loads and medium level vehicle dynamics responses can be accurately predicted but also low-level (on-center performance) lateral accelerations. A detailed discussion is provided in [69].

An alternative data-driven approach was evaluated for the case of repetitive vehicle durability testing. One (or a limited set) of vehicles were instrumented with advanced (expensive and time-consuming to instrument) wheel load (force and torque) sensors as well as a set of standard sensors (accelerometer, IMU…). A data-driven Executable Digital Twin was developed through training of a Neural Network with road test data. The Neural Network was then exported (e.g., in ONXX format) and used in subsequent tests on the same as well as similar vehicles only using the basic sensor set. Figure 10 illustrates this process.

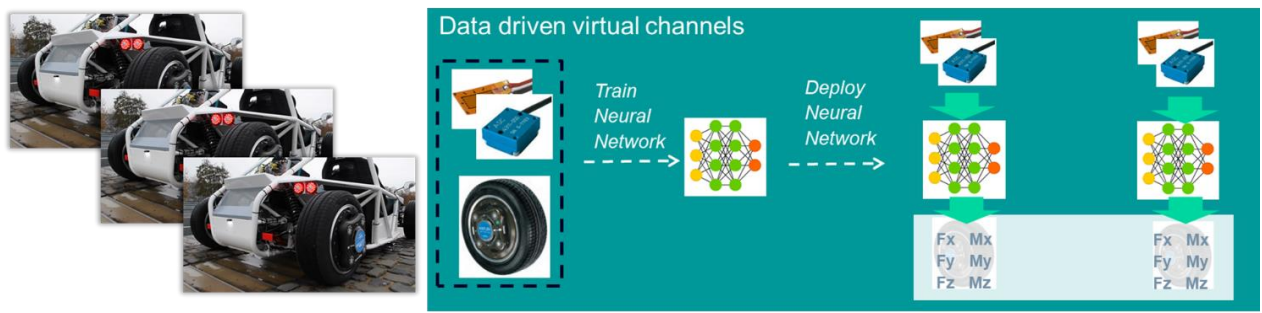

Figure 10: Vehicle axle load estimation using a data driven approach.

## 5.2 Other Executable Digital Twin use cases

Virtual sensing constitutes at this point probably the best documented use scenario for the Executable Digital Twin in our application fields. Many other cases are realized or are under development, covering a wide range of domains, from design to manufacturing and process control (e.g. see [70]).

Besides virtual sensing other Executable Digital Twin cases are increasingly documented in relation to Hardware-in-the-loop, controls development and manufacturing and process control.

### 5.2.1 Hardware-in-the-loop drivetrain analysis

Executable Digital Twin models are well suited for hybrid testing applications where part of the system under test is physical and part is virtual. Application cases are numerous, for example to test and evaluate the integration of new (physically available) components in a virtual system and/or optimize their performance without executing the physical integration. Alternatively, control systems, controller strategies or even embedded control software can be developed and tuned without that the full system to be controlled must be physically present.

The present case addresses the requirement to develop and optimize novel electric powertrains suited for a broad range of user needs and driving scenarios. The demands for high levels of energy efficiency, performance and comfort leads to ever more integrated powertrains and sophisticated controllers that must be evaluated for various vehicle application platforms and driving scenarios. This requires frontloading the validation and optimization testing to the subsystem level, replicating full vehicle integration testing. A powerful new XiL setup (X-in the Loop, with X standing for Hardware, Software, Driver…) for electric drivelines was developed hereto. Real-time Executable digital Twin models can represent the virtual system components, the vehicle to be integrated into as well as the driving loading conditions (such as e.g., regenerative braking) The investigated performances can include energy consumption, thermal behavior, transient responses s relevant for performance, driveability and comfort. The impact of design changes on the vehicle level on the powertrain performance can be assessed, such as modified battery packs, gearbox ratios etc. A detailed discussion is provided in [71]. Figure 11 presents an overview of the XiL test setup.

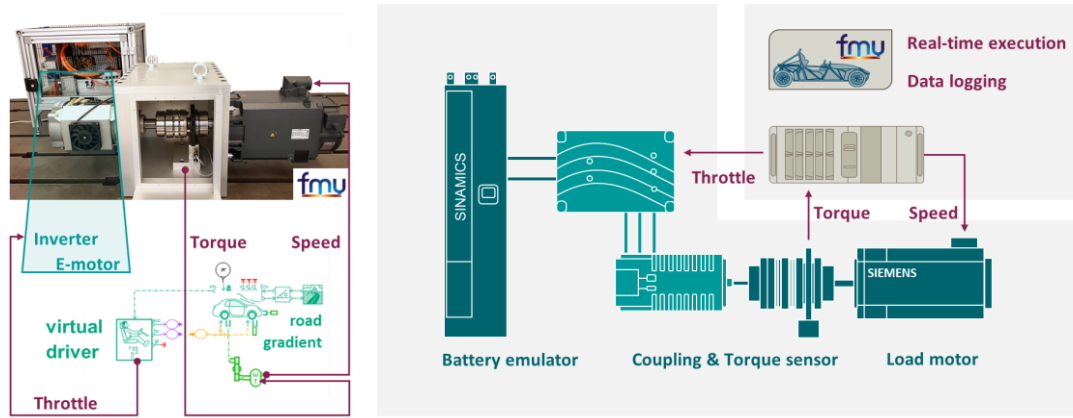

Figure 11: HiL setup with e-motor and invertor. Figure courtesy Forrier et al. [71]

### 5.2.2 Fail-safe testing of in-wheel drivetrain

Also, the potential to evaluate fault scenario's or checking fail-safe operation without having the full system available is of large interest to optimizing the validation and verification process.

The presented case addresses testing of failure modes of an in-wheel motor (IWM). The full-vehicle response is critical as one failing motor must be compensated by the other IWM for vehicle safety. The device-under-test is the Propulsion Control Unit (PCU) which controls the four in-wheel motors of the vehicle. Various fault scenarios are investigated (short circuit, loss of communication, torque inversion…) for one of the motors. However, injecting critical faults in the system-under-test is dangerous and expensive to test in a full electric vehicle. A Hardware-in-the-Loop testbench allows to evaluate the full vehicle behavior in a safe way using the hardware of the motor under test (M1), controlled over CAN bus from the real-time platform, and Executable Digital Twin models for the vehicle (vehicle dynamics, driveline) and the three other IWM driving the actuation motor (M2). This xDT model runs on the real-time platform of the testbench. Figure 12 shows the test setup. A detailed discussion is available in [72]

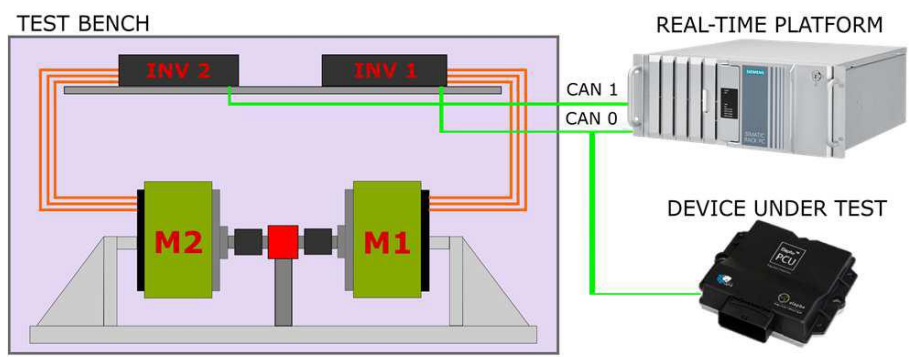

Figure 12: In-wheel real-time test bench configuration. Figure courtesy Sputh et al. [72]

### 5.2.3 Manufacturing control

Models that are used in manufacturing control or robotics are often developed ad-hoc, not derived from the digital twin established in the design engineering phase. The Executable digital Twin hence offers significant potential to leverage the digital twin to the manufacturing execution level. Examples are in virtual commissioning, in virtual sensing (e.g., when linked to the sensors available on the Industrial Edge or PLC) but also to support model-based process control.

An example is robot milling. The accuracy of milling machines is typically limited by the mechanical stiffness of the corresponding machine. Process forces of several hundred Newton led to deformations of the machine affecting the accuracy of the produced part, e.g., in the context of standard industrial robots, milling process forces could lead to deformations in the range millimeters, which is well above most industrial requirements. Combining first principal predictions of process forces with a mechanical robot model and online calibration technologies allows the prediction of expected deflections of the robot and to compensate them correspondingly in the control cycle, with update rates on the order of 5 ms [73]. This results in the reduction of machining errors of robots by 90%, sufficient for milling tasks. Classical machine learning approaches will not work since on the one hand metrology in milling machines is very limited (due to dirt, splinter, etc.) and on the other hand combinatorial options in terms of geometries, materials, milling paths, and robot poses would not allow to sample sufficient data. Figure 13 shows a typical result.

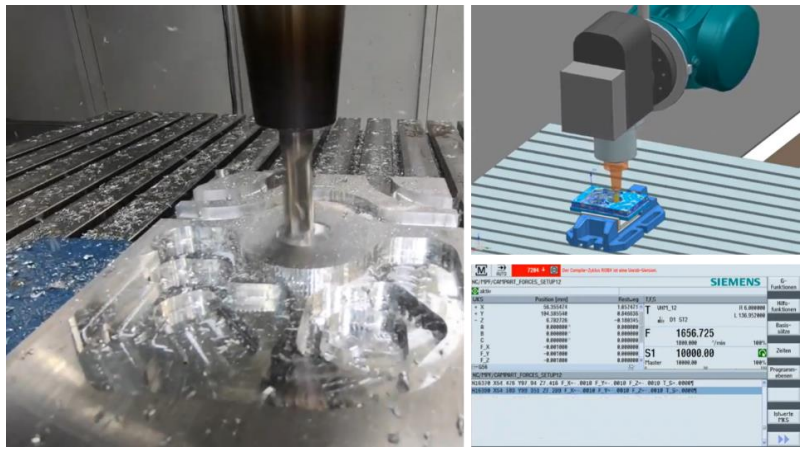

Figure 13: Digital Twin-based control solution increasing accuracy of milling robots as a key enabler for industrial metal milling.

### 5.2.4 Model Predictive Control

While using models for controller design is common practice, the use of the models in the control, in model-based control is less widespread. The used models are primarily made ad-hoc. A large potential exists for deriving such models from the digital twin that is in most cases anyway developed during the design and engineering stages. An application case was realized, using the available calibrated vehicle dynamics (digital twin) model to design a nonlinear model predictive controller for autonomous driving [74].

## 6 Conclusions and Outlook

The Executable Digital Twin is a logical consequence of further integrating the digital and physical world using the digital twin paradigm. By generating from the digital twin, a dedicated and executable model for a specific execution environment and connecting this to the physical asset, a series of powerful applications are brought to a next level of effectiveness. Deployments of the Executable Digital Twin for virtual sensing,

X-in-the-loop, model predictive control, and model-based diagnostics bring these applications from an ad-hoc trial-and-error level to an integrated part of the lifecycle engineering process.

First use cases show the intrinsic potential of the approach in full product life cycle, from design and design validation, to manufacturing, commissioning, and system operation. The key element is that by closing the loop between the physical and the digital world, the digital twin approach can show its full potential.

The enablement of the Executable Digital Twin however relies strongly on several powerful technology building blocks such as Model Order Reduction, State Estimation, and Co-Simulation to which it poses novel challenges and hence for which dedicated further research efforts will be needed. Research for the next generations of the Executable Digital Twin will include bringing the models to the level of hardware implementation (FPGA, custom IC) [75], hybrid co-simulation across heterogenous (and physically non-collocated) platforms, enabled by 5G and 6G communication networks, as well novel modeling and simulation paradigms. As the Executable Digital Twin allows for large scale deployment and even trading of models, the concept is expected to enable new business models and might be a potential building block for the Metaverse [76].

## Acknowledgements

The presented work reflects a joined research endeavor from teams at Siemens Digital Industries Software (Simcenter product line), Siemens Technology (Simulation and Digital Twin Technology Field), and KU Leuven and Flanders Make as strategic research partners. The research is supported by multiple projects funded by the European Commission, Vlaio (Flanders Innovation & Entrepreneurship Agency) and the KU Leuven Industrial Research Fund and constitutes a key part of the Siemens Company Core Technology Simulation and Digital Twin, the support of all of which is gratefully acknowledged.